\documentclass[usenatbib]{mn2e}
\usepackage{xspace}
\usepackage{amsmath}
\usepackage{epsfig}
\usepackage{defs}

\title[Dark Matter Haloes and Their Primordial Antecedents]{The Formation of Spiral Galaxies: Adiabatic Compression with Young's Algorithm and the Relation of Dark Matter Haloes to Their Primordial Antecedents}

\author[H. Katz et al. ]{Harley Katz$^{1,2}$\thanks{E-mail: hk380@ast.cam.ac.uk}, Stacy S. McGaugh$^{3}$, J. A. Sellwood$^{4}$, and W. J. G. de Blok$^{5,6,7}$\\
$^1$Department of Astronomy, University of Maryland, College Park, MD 20742, USA\\
$^2$Institute of Astronomy, University of Cambridge, Madingly Road, Cambridge, CB3 0HA\\
$^3$Department of Astronomy, Case Western Reserve University, Cleveland, OH 44106, USA\\
$^4$Rutgers University, Department of Physics \& Astronomy, 136 Frelinghuysen Road, Piscataway, NJ 08854, USA\\
$^5$Netherlands Institute for Radio Astronomy (ASTRON), Postbus 2, 7990 AA Dwingeloo, the Netherlands\\
$^6$Astrophysics, Cosmology and Gravity Centre, Department of Astronomy, University of Cape Town, Private Bag X3,\\ Rondebosch 7701, South Africa\\
$^7$Kapteyn Astronomical Institute, University of Groningen, PO Box 800, 9700 AV Groningen, the Netherlands\\
}

\begin{document}

\maketitle

\begin{abstract}
We utilize Young's algorithm to model the adiabatic compression of the dark matter haloes of galaxies in the THINGS survey to determine the relationship between the halo fit to the rotation curve and the corresponding primordial halo prior to compression.  Young's algorithm conserves radial action and angular momentum, resulting in less halo compression than more widely utilized approximations.  We find that estimates of the parameters of NFW haloes fit to the current dark matter distribution systematically overestimate the concentration and underestimate the virial velocity of the corresponding primordial halo.  It is the latter that is predicted by dark matter simulations; so accounting for compression is a necessary step for evaluating whether massive galaxies are consistent with dark matter-only simulations.  The inferred primordial haloes broadly follow the c-$V_{200}$ relation expected in a $\Lambda$CDM cosmogony, but often scatter to lower concentrations.  We are unable to obtain fits at all for those galaxies whose current dark matter haloes are poorly described by the NFW form.  We thus find a mixed bag:  some galaxies are reasonably well described by adiabatic compression within a primordial NFW halo, while others require an additional mechanism that reduces the density of dark matter below the primordial initial condition.
\end{abstract}


\section{Introduction}
In the early Universe, the dark matter and baryons are well mixed \citep{Spergel2003}.  However, as the baryons cool and dissipate energy, they fall to the center of the dark matter haloes and form the various types of galaxies \citep{White1978,Fall1980,Gunn1982}.  The dark matter halo responds gravitationally to the infall and settling of baryons by compressing.  Initially considered as an adiabatic process in an isolated dark matter halo, this same process also pertains in the context of the modern hierarchical picture of galaxy formation \citep{Choi2006}.  The compression of dark matter haloes by baryonic infall is an inevitable consequence of the rearrangement of mass necessary to form a galaxy.  It distorts the initial NFW density profiles predicted by cosmological structure formation simulations \citep{NFW1997} to a form that lacks an analytical description.

The Blumenthal method \citep{Blumenthal1986} has been the standard in the literature since its initial development; however, multiple groups have shown that this formalism tends to over predict the compression of the halo (e.g. \cite{Barnes1987,Sellwood1999,Gnedin2004}).  The drawback of this method is that it assumes all particles in the halo are on circular orbits (which is equivalent to assuming that the radial action of all such particles is zero).  Historical examples of computing adiabatic compression exactly have required computationally expensive hydrodynamic simulations of cooling gas \citep{Gottloeber2002, Abadi2003, Governato2004}.  However, \cite{Young1980} developed a method to compute the adiabatic compression of spherical systems exactly by also accounting for random motions of particles in the halo (thus conserving the radial action in addition to the angular momentum).  This method was first applied to the compression of dark matter haloes by \cite{Wilson2004} and \cite{Sellwood2005}.  Disc galaxies are not spherical, but the monopole term dominates so that Young's method is as accurate as the equivalent (and computationally far more expensive) fully hydrodynamic simulation \citep{Sellwood2005}.

The Blumenthal adiabatic contraction method predicts steeper inner profiles for initially NFW haloes than observed, which led \cite{Dutton2013} to suggest that haloes remain uncontracted or even expand slightly during galaxy formation. Adiabatic contraction cannot simply be switched off, so some additional mechanism must be invoked to mitigate its effects.  A partial solution may be offered by light-weight IMFs; by reducing the mass in stars we can reduce the amount of compression they induce.  However, the small scatter in the Tully-Fisher relation precludes arbitrary adjustment of the IMF, while its normalization is consistent with high surface brightness discs being nearly maximal \citep{mcgaugh2005} like the Milky Way \citep{flynn,bovyrix}.  Because Young's algorithm correctly predicts less contraction than the Blumenthal algorithm, it allows for an IMF that is heavy enough to be consistent with these results.

Adiabatic contraction is not the only process to dynamically shape the dark matter halo of a galaxy.  Feedback driven outflows \citep{Navarro1996,Read2005,Pontzen2012} and dynamical friction \citep{Elzant2001,Weinberg2002,Johansson2009} can in principle counteract the compression of the halo.  However, \cite{Sellwood2008} have shown that the bar-halo friction proposed by \cite{Weinberg2002} cannot cause a significant reduction in density (see also \cite{McMillan2005}).  The mechanism described by \cite{Elzant2001} likely results in an insignificant density reduction due to a number of issues described by \cite{Jardel2009}, and an extremely massive satellite ($\sim1\%$ of the virial mass of the halo) would have to sink in the galactic potential in order to cause a significant density reduction \cite{Sellwood2013}.  The magnitude we infer for such mechanisms as feedback driven outflows and dynamical friction depends on the amount of adiabatic contraction that occurs in the first place. In order to constrain the strength of feedback mechanisms, one must first correctly model the compression.   

Here, we extend the study of \cite{Sellwood2005} by applying Young's algorithm to a sample of galaxies from the THINGS \citep{Walter2008,dB2008} survey and look to relate the observed properties of galaxies with the characteristics of the primordial halo.  In Section 2, we present our results in applying the algorithm to galaxies from the THINGS survey.  In Section 3, we develop an analytical method for relating observed parameters of spiral galaxies to their primordial NFW halo and compare the predictions with observational constraints from lensing surveys.  In Section 4, we summarize our conclusions.  

\section{Data and Analysis}
\subsection{The THINGS Sample}
We adopt as our sample the THINGS galaxies analyzed by \cite{dB2008}.  This provides a sample of well resolved disc and gas-rich dwarf galaxies with observations of both stars and gas, both of which are essential to proper mass modeling.  The stellar mass model is provided by Spitzer $3.6\mu$ observations while the atomic gas content has been observed with the VLA \citep{Walter2008}.  

\cite{dB2008} fit NFW models to the data for definite prescriptions for the stellar mass-to-light ratio provided by population synthesis models.  These fits to the \textit{current} distribution of dark matter provide a reference point for interpreting the effects of adiabatic contraction.  In effect, we obtain the ``before and after'' picture of the dark matter distribution given the assumption of a primordial NFW halo that has adiabatically contracted under the influence of the observed distribution of stars and gas.

As discussed above, these assumptions provide a vital starting point for relating the results of structure formation simulations with real observed galaxies.  These assumptions may not be sufficient in all cases.  It is therefore important to understand what we do not fit as well as what we do fit.

We have applied Young's method to 12 of the 17 galaxies modeled by \cite{dB2008}.  Of the 12 galaxies modeled, the fits found are generally acceptable, modulo the expected minor issues discussed in detail below.  Of the five remaining galaxies, one is NGC 6946.  This galaxies is rather face-on, and the resulting mass model is  sensitive to debatable choices for the inclination.  We therefore omit it from consideration.  

Four galaxies, NGC 925, NGC 2366, NGC 2976, and IC 2574, cannot be fit with an NFW halo.  \cite{dB2008} find unphysically low concentrations for the current dark matter distribution.  This is the classic sign of galaxies that suffer from the cusp-core problem \citep{dBMR2001,KZN2008,Oh2011}.  Adiabatic contraction tends to increase the concentration of the primordial dark matter halo, so an unphysically low concentration in the \textit{current} dark matter halo implies an antecedent \textit{primordial} halo whose concentration is still less consistent with $\Lambda$CDM.  These are galaxies for which feedback or some other mechanism must be invoked in order to rearrange the primordial halo mass distribution predicted by structure formation simulations.  

Modeling feedback is beyond the scope of this paper, and the many attempts to do so are still far from satisfying observational requirements \citep{McGaugh2004}.  Rather than attempt to model every conceivable effect, our aim is to simplify the problem in order to gain physical insight into one basic process of galaxy formation. We focus on adiabatic contraction because we know that it is a physical effect that must be important, and which largely precedes any subsequent feedback from star formation or AGN in the condensed baryonic mass.  It is necessary to get this step right before determining the magnitude of required feedback effects.

We should not read too much into the numbers of galaxies that can and cannot be fit.  The THINGS project provides ideal data for our exercise.  It includes a nice spectrum of disc galaxy properties, but it does not provide a complete sample that is representative of the numbers of galaxies at each property.  We shall see that of the galaxies that can be fit, the inferred primordial haloes are often consistent with the concentration--virial mass relation predicted by $\Lambda$CDM.  Sometimes the data fall to lower concentrations than expected, while it is rare for the inferred concentration to run high.  Among the galaxies that we can not fit, the concentrations are all too low.  This is the general rule for low mass and low surface brightness galaxies \citep{dB2010}.  This should be kept in mind when interpreting the distribution of fit parameters for those galaxies that can be fit.

\subsection{Application of Young's Method}
We assume that the primordial haloes exhibit spherical, NFW potentials of the form
\begin{equation}
\Phi_{NFW}(r)=-\frac{GM_s}{r_s}\frac{\ln(1+(r/r_s))}{r/r_s}.
\end{equation}
Here $M_s=4\pi\rho_s r_s^3$ and $\rho_s=4\rho(r_s)$.  These can be related to the more frequently used concentration $c$, and virial velocity, $V_{200}$ by noting that $c=r_{200}/r_s$ and $V_{200}=10cr_sH_0$.

The two parameter NFW profile has been superseded by the three parameter Einasto profile \citep{Nav2004,Merritt2005} in fitting simulated haloes (see \citealt{Chemin2011} for Einasto fits to the THINGS sample).  While we could implement an Einasto profile, we choose to use the NFW profile for several reasons.  First, we wish to minimize the number of free parameters.  Second, the difference between NFW and Einasto profiles is not observationally distinguishable \citep{McGaugh2007}, so there is no added value in attempting to make this distinction.  Finally, the relation between NFW parameters and the parameters of the cosmology in which haloes form is well documented \citep{MBdB2003,Maccio2008}.

Even limiting ourselves to the two parameter NFW model, we expect some degeneracy among our fit parameters.  There is a strong degeneracy between stellar mass and halo parameters \citep{Kent1987} stemming from the fact that a single parameter suffices to describe rotation curve data \citep{McGaugh2004}.  Even with a fixed stellar mass as implemented here, NFW haloes are largely self-degenerate over the finite range of radii constrained by real observational data: one NFW model looks much like another at small radii even if they differ greatly at large radii.  Nonetheless, we can explore $M_s-r_s$ space and obtain a best fit.

\begin{figure}
\epsfig{figure=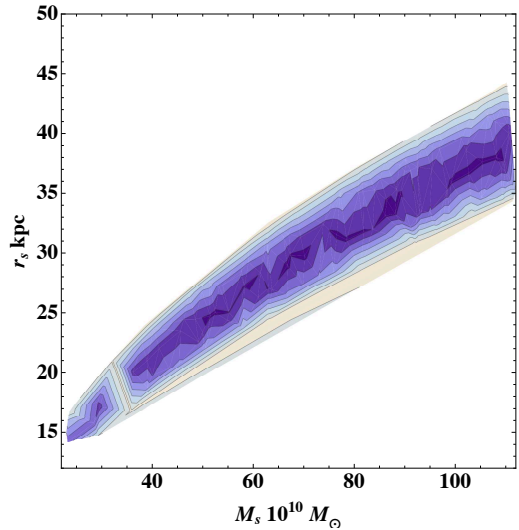,scale=0.35,trim= 25 100 100 100}
\caption{Example of the initial coarse grid for NGC 2403.  Contours represent regions of constant $\chi^2$.  The crescent shape of this parameter space is consistent over all galaxies.}
\label{cg} 
\end{figure}

For each galaxy, we initially define a large range in $M_s$ and $r_s$, and build a Monte Carlo grid to explore this parameter space.  After these initial simulations, we further refine the grid about the best fit parameters to find the minimum $\chi^2$.  In Figure \ref{cg}, we show example of a contour plot for $\chi^2$ of initial coarse grid in the $M_s-r_s$ parameter space of NGC 2403.  The crescent shape in the $M_s-r_s$ parameter space is consistent for all galaxies in our sample, and is typical of the covariance of NFW fit parameters \citep{dBMR2001}.

We emphasize that the error bars on the rotation curves are only estimates of the true uncertainties which attempt to take into account error in the tilted ring fits, uncertainties in inclination, differences in the velocities of the approaching an receding sides of the galaxy as well as other effects (for a more detailed description of error bars, refer to \citealt{dB2008}).  While this all-inclusive approach is likely an overestimate of the true error, the uncertainty in the assumption of IMF is likely to dominate over the uncertainties in the rotation curve.  Thus the measured values of $\chi^2$ are unlikely to be completely representative of the true accuracy of the fit and are rather used as mechanisms to fairly derive compressed halo fits in a comparable way to the uncompressed NFW haloes of \cite{dB2008}.  

The degree of compression of the primordial halo is directly related to the chosen stellar IMF.  Unfortunately, the mass-to light ratios for these galaxies are reasonably unconstrained.  \cite{dB2008} provides three NFW fits using a diet-Salpeter IMF (a version of the Salpeter IMF scaled down in mass by 70\% to not exceed maximum disk as estimated by \citealt{Bell2001}), a Kroupa IMF \citep{Kroupa2001}, and a free $M_*/L$ (where the change in $M_*/L$ as a function of radius is given by the color gradient and the resulting stellar rotation curve is then scaled with a uniform constant). The Kroupa IMF has a flatter slope at $M<0.5M_{\odot}$ and so is less massive than the diet-Salpeter IMF.  We therefore expect less compression when adopting the Kroupa IMF than with the diet-Salpeter IMF.


For our study, we have chosen to model all galaxies using the diet-Salpeter IMF, understanding that this may not represent the exact $M_*/L$ for all galaxies in the THINGS survey.  A Salpeter-like IMF is favored in the spiral bulge models of \cite{Dutton2013} which serves as additional motivation for this choice of IMF.  Furthermore, the choice of the diet-Salpeter IMF demonstrates how Young's algorithm can work with a heavier IMF without over predicting compression.  We provide three examples of what we denote as ``Low Concentration Galaxies" (LCGs), where the baryonic component is super maximal for this choice of IMF, causing the NFW fit to have an extremely low concentration.  We provide an additional fit for these three galaxies, assuming the Kroupa IMF\footnote{What \cite{Bell2001} call a Kroupa IMF does not include brown dwarfs.  Including these, as would be appropriate here since we are concerned with the entire baryonic mass of the stellar disc, would result in a mass similar to the diet-Salpeter IMF.  We maintain the distinction to illustrate the effects of the uncertainty in the IMF.} to demonstrate how the $M_*/L$ can play a significant role in the rotation curve fit.  One expects a fair amount of scatter in the relation between color and mass-to-light ratio for any given IMF, so it is likely that the LCGs are simply galaxies that happen to scatter low in $M_*/L$ rather than having intrinsically different IMFs.  Nevertheless, one should bear in mind that the absolute normalization of the IMF is an irreducible systematic uncertainty in this (or any similar) exercise.

Table 1 lists  the galaxies used in our study.  This includes specific components for each model.  Table 2 lists parameters for the LCGs which have been refitted using a Kroupa IMF.  For the remainder of this section, we will refer to the uncompressed primordial dark matter halo as ``PH", the compressed dark matter halo as ``CH", and the NFW halo which \cite{dB2008} fits to each rotation curve as ``dBH".

\begin{table*}
\label{tabone}
\centering
\begin{tabular}{@{}lcccccccccccc@{}}
\multicolumn{12}{c}{Mass Models with Fixed $\Upsilon_*^{3.6}$ and Diet-Salpeter IMF}\\
\hline
\hline
 &  &  &  &  & \multicolumn{3}{c}{de Blok Parameters} & & \multicolumn{3}{c}{This Paper} \\
 \cline{6-8} \cline{10-12} \\ 
Galaxy & log$M_*^D$ & $\Upsilon_{*,D}^{3.6}$ & log$M_*^B$ & $\Upsilon_{*,B}^{3.6}$  & $c$ & $V_{200}$ & $\chi_{\nu}^2$ & & $c$ & $V_{200}$ & $\chi_{\nu}^2$\\
(1) & (2) & (3) & (4) & (5) & (6) & (7) & (8) & & (9) & (10) & (11)\\
\hline
NGC 2403 (1 comp) & 9.71 & 0.41 & ... & ... & $9.9\pm0.2$ & $109.5\pm1.0$ & 0.55 & & 6.2 & 134.9 & 0.64 \\
NGC 2403 (2 comp) & 9.67 & 0.39 & 8.63 & 0.60 & $9.8\pm0.2$ & $110.2\pm1.0$ & 0.56 & & 6.1 & 140.0 & 0.68 \\
NGC 2841 & 11.04 & 0.74 & 10.40 & 0.84 & $16.1\pm0.2$ & $183.2\pm1.2$ & 0.42 & & 10.2 & 193.8 & 0.36 \\
NGC 2903 (outer) & 10.15 & 0.61 & 9.33 & 1.30 & $30.9\pm0.6$ & $112.9\pm0.6$ & 0.36 & & 21.2 & 118.7 & 0.32 \\
NGC 3031 & 10.84 & 0.80 & 10.11 & 1.00 & $3.0\pm2.9$ & $190.9\pm161.1$ & 4.36 & & 2.4 & 187.6 & 4.40 \\
NGC 3198 (1 comp) & 10.40 & 0.80 & ... & ... & $7.5\pm0.4$ & $112.4\pm2.1$ & 1.37 & & 3.5 & 129.4 & 1.83 \\
NGC 3198 (2 comp) & 10.45 & 0.80 & 9.46 & 0.73 & $5.1\pm0.5$ & $122.7\pm4.9$ & 2.88 & & 3.1 & 147.0 & 2.94 \\
NGC 3621 & 10.29 & 0.59 & ... & ... & $3.7\pm0.2$ & $165.5\pm5.9$ & 0.81 & & 3.2 & 174.1 & 1.84 \\
NGC 4736 & 10.27 & 0.63 & $9.59^i$ & $0.33^i$ & $11.4\pm9.8$ & $35.2\pm0.3$ & 1.51 & & 7.6 & 39.8 & 1.74 \\
DDO 154 & 7.42 & 0.32 & ... & ... & $4.4\pm0.4$ & $58.7\pm4.3$ & 0.82 & & 4.5 & 72.4 & 1.11 \\
NGC 7793 & 9.44 & 0.31 & ... & ... & $5.8\pm1.4$ & $156.6\pm39.1$ & 4.17 & & 4.4 & 188.7 & 5.11\\
 \multicolumn{12}{c}{\underline{Low Concentration Galaxies}}\\
 NGC 3521 & 11.09 & 0.73 & ... & ... & $<0.1$ & $403.2\pm123.2$ & 8.52 & & 1.6 & 201.6 & 11.31\\
 NGC 5055 & 11.09 & 0.79 & $9.32^{ii}$ & $0.11^{ii}$ & $<0.1$ & $450.1\pm32.4$ & 10.31 & & 1.6 & 163.6 & 15.20\\
 NGC 7331 (const)$^{iii}$ & 11.22 & 0.70 & 10.24 & 1.00 & $<0.1$ & $>500$ & 4.08 & & 1.5 & 217.3 & 4.64\\
 \hline
\end{tabular}
\caption{$i.$ $M_*^B$ and $\Upsilon_{*,B}^{3.6}$ were left as free parameters and differ from predictions (see \protect\cite{dB2008} Table 4).
 $ii.$ The mass model neglects the colour gradient (see \protect\cite{dB2008} Table 4).}
\end{table*}

\begin{table*}
\label{tab:two}
\centering
\begin{tabular}{@{}lcccccccccccc@{}}
\multicolumn{12}{c}{Mass Models with Fixed $\Upsilon_*^{3.6}$ and Kroupa IMF}\\
\hline
\hline
 &  &  &  &  & \multicolumn{3}{c}{de Blok Parameters} & & \multicolumn{3}{c}{This Paper} \\
 \cline{6-8} \cline{10-12} \\ 
Galaxy & log$M_*^D$ & $\Upsilon_{*,D}^{3.6}$ & log$M_*^B$ & $\Upsilon_{*,B}^{3.6}$  & $c$ & $V_{200}$ & $\chi_{\nu}^2$ & & $c$ & $V_{200}$ & $\chi_{\nu}^2$\\
(1) & (2) & (3) & (4) & (5) & (6) & (7) & (8) & & (9) & (10) & (11)\\
\hline
 NGC 3521 & 10.94 & 0.52 & ... & ... & $8.9\pm2.0$ & $128.4\pm16.4$ & 5.55 & & 3.8 & 102.1 & 5.27\\
 NGC 5055 & 10.94 & 0.56 & $9.81^{i}$ & $0.34^{i}$ & $2.1\pm0.4$ & $217.8\pm21.2$ & 1.45 & & 1.9 & 212.7 & 2.83\\
 NGC 7331 (const)$^{ii}$ & 11.07 & 0.50 & 10.09 & 0.71 & $4.9\pm0.4$ & $200.0\pm10.7$ & 0.24 & & 3.7 & 210.9 & 0.30\\
 \hline
\end{tabular}
\caption{ $i.$ $M_*^B$ and $\Upsilon_{*,B}^{3.6}$ were left as free parameters and differ from predictions (see \protect\cite{dB2008} Table 5).
 $ii.$ The mass model neglects the colour gradient (see \protect\cite{dB2008} Table 5).} 
\end{table*}

\subsection{Fits for Individual Galaxies}
\subsubsection{NGC 2403}
\begin{figure}
\epsfig{figure=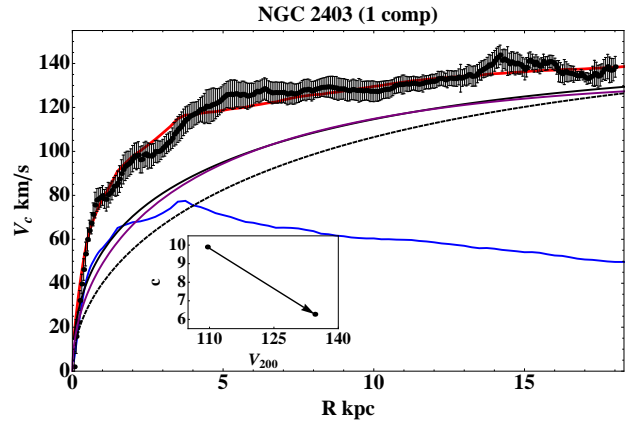,scale=0.37}
\caption{Rotation curve fit for NGC 2403 one component model.  The blue line represents the total baryonic component (stars $+$ gas).  The purple line shows the dBH.  The dashed black line is the PH and the solid black line is the CH.  Finally, the red line is the fit from summing the CH (solid, black) with the baryonic component (blue).  This inset shows the best fit c and $V_{200}$ for the dBH with an arrow pointing to the best fit parameters for the PH as calculated in this paper.}
\label{NGC24031} 
\end{figure}

\begin{figure}
\epsfig{figure=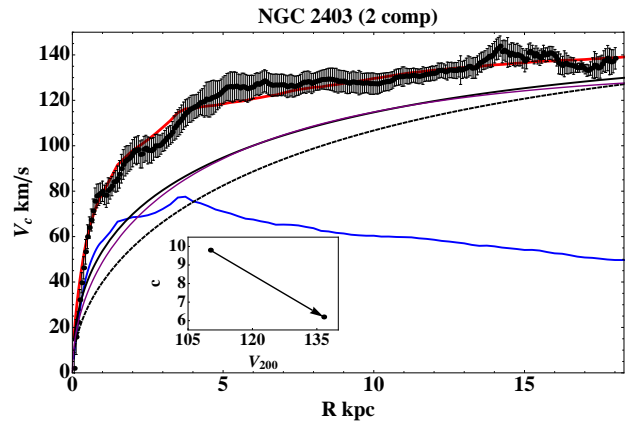,scale=0.37}
\caption{Rotation curve fit for NGC 2403 two component model.  Lines and inset are the same as in Figure \protect\ref{NGC24031}.}
\label{NGC24032} 
\end{figure}

NGC 2403 is a high surface brightness late-type Sc spiral galaxy.  We have fit two models for this galaxy, the first assuming a one component disk, and the second assuming an additional bulge component.  Figure \ref{NGC24031} depicts the former while Figure \ref{NGC24032} depicts the latter.  We can see that they are nearly identical.  Our values for $\chi_{\nu}^2$ for both CH fits are very comparable to the dBHs.  As expected, our CHs have slightly more mass within $\sim4$ kpc where the baryonic component of the rotation curve is approximately equal to that of the dark matter.  Our CH is indistinguishable from the dBH out to $\sim15$ kpc where our CH continues to rise.

For the one component disk model, our PH has a $23\%$ higher $V_{200}$ than that of the dBH while the concentration has decreased by a factor of $64\%$.  Similarly, the PH of the two component model has a $V_{200}$ that is $23\%$ higher than that of the dBH while the concentration has decreased by a factor of $63\%$.

\subsubsection{NGC 2841}
\begin{figure}
\epsfig{figure=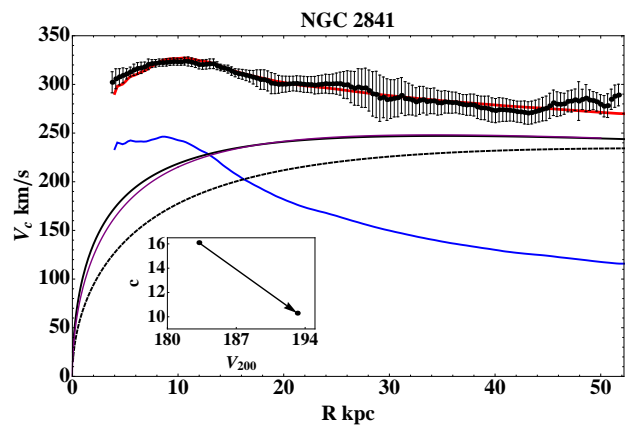,scale=0.37}
\caption{Rotation curve fit for NGC 2841.  Lines and inset are the same as in Figure \protect\ref{NGC24031}.}
\label{NGC2841} 
\end{figure}

NGC 2841 is an early-type (Sb) spiral galaxy.  We use a two component model to represent the bulge and disk however the rotation curve data does not represent the inner 3 kpc.  Our CH fits the rotation curve quite well (see Figure \ref{NGC2841}).  The CH is nearly identical to the dBH at all radii; however as expected, the CH has slightly more mass within $\sim13$ kpc.  The $V_{200}$ of our PH is only $5.5\%$ larger than that of the dBH and the concentration has been reduced by a factor of $64\%$.

\subsubsection{NGC 2903}
Here we only study the outer portion of the rotation curve of NGC 2903, an SBd galaxy, due to noncircular motions towards the center.  We use a two component model to represent the bulge and disk and notice a slight decline in the rotation curve at large radii as well as an interesting feature around 10 kpc.  This small jump is likely due to the titled ring fit which demonstrates an abrupt shift in inclination at this radius.  The inclination, however, is well behaved at all radii further out.  

\begin{figure}
\epsfig{figure=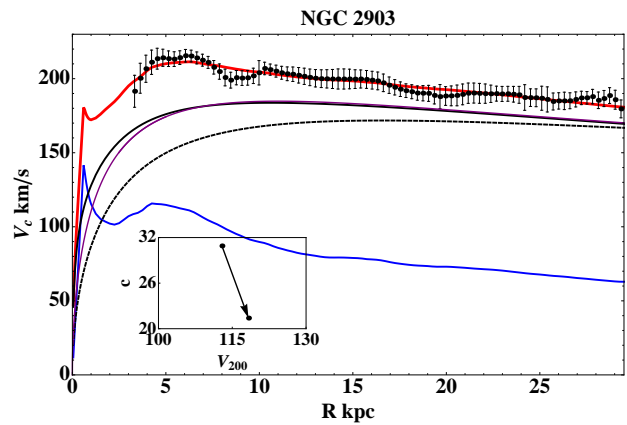,scale=0.37}
\caption{Rotation curve fit for NGC 2903.  Lines and inset are the same as in Figure \protect\ref{NGC24031}.  The innermost points have been omitted from the plot.}
\label{NGC2903} 
\end{figure}

Our CH the rotation curve very well (see Figure \ref{NGC2903}).  Our fit is indistinguishable from the dBH for all radii greater than about 6 kpc, and again, the CH has slightly more mass in the inner parts.  The $V_{200}$ of our PH is only $5.1\%$ larger than that of the dBH and the concentration has been reduced by a factor of $61\%$.  

\subsubsection{NGC 3031}
NGC 3031 is a grand design spiral galaxy.  The baryonic contribution to the rotation curve is modeled by a two component disk which is maximal.  The baryonic component is completely dominant over the dark matter for all sampled radii.  Features in the rotation curve, including the large bump spanning from $R=6-8$ kpc is likely due to an abrupt shift in inclination caused by nonuniform circular motion.  The inclination increases steadily after this feature which might result in the decline is the rotation curve at large radii.

\begin{figure}
\epsfig{figure=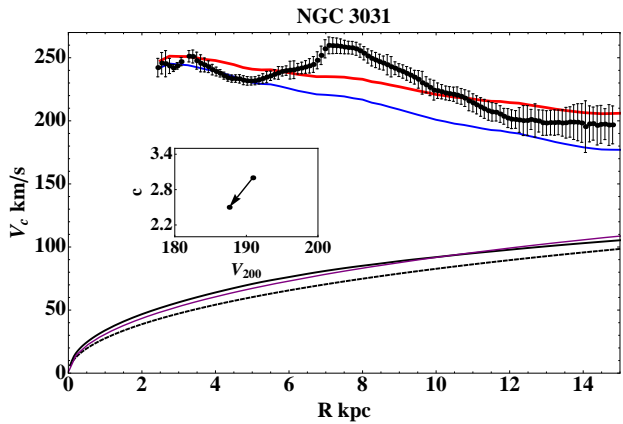,scale=0.37}
\caption{Rotation curve fit for NGC 3031.  Lines and inset are the same as in Figure \protect\ref{NGC24031}.}
\label{NGC3031} 
\end{figure}

No smooth dark matter halo will be able to fit this rotation curve well due to the large bump at $6<R<8$ kpc because the baryonic component is not only dominant, but also smooth.  Thus while our CH fit is comparable the dBH, our $\chi^2$ calculation arrives at this value by slightly over predicting the inner and outer region while under predicting the aforementioned feature in the rotation curve (see Figure \ref{NGC3031}).

\subsubsection{NGC 3198}
NGC 3198 is an SBc galaxy.  Similar to NGC 2403, we study both a one and two component model for the disk.  In both models, the baryonic component alone over predicts the rotation curve at $R\sim1$ kpc; however, this is less extreme in the one component disk model.  \cite{dB2008} note that there is evidence for the presence of a small bar which could affect the inner part of the rotation curve due to noncircular motions and cause the baryonic component to over predict the rotation curve.  

\begin{figure}
\epsfig{figure=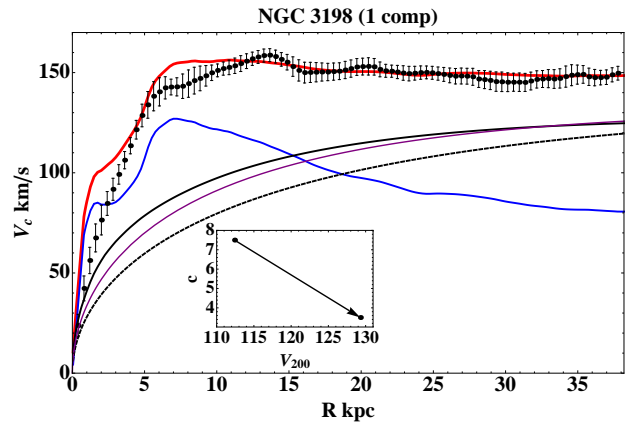,scale=0.37}
\caption{Rotation curve fit for NGC 3198 one component model.  Lines and inset are the same as in Figure \protect\ref{NGC24031}.}
\label{NGC3198a} 
\end{figure}

\begin{figure}
\epsfig{figure=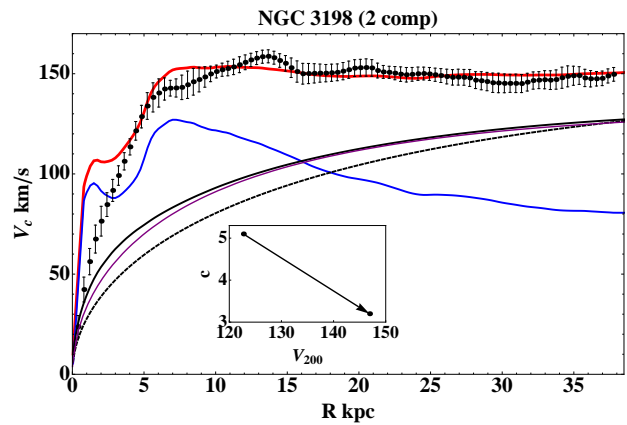,scale=0.37}
\caption{Rotation curve fit for NGC 3198 two component model.  Lines and inset are the same as in Figure \protect\ref{NGC24031}.}
\label{NGC3198b} 
\end{figure}

Our CHs fit the outer portions of the rotation curve well but the value for $\chi_{\nu}^2$ is plagued by the over prediction of the rotation curve by the baryonic component.  Our fit for the two component model is very comparable to the dBH for the same model with very slightly more mass at radii less than $\sim15$ kpc (see Figure \ref{NGC3198b}).  However, the CH model for one component disk is not as good as the dBH, which has less mass at all radii out to $R\sim30$ kpc (see Figure \ref{NGC3198a}).  Unlike NGC 2403, the one and two component models for NGC 3198 result is reasonably different PHs.  For the one component disk model, the $V_{200}$ of our PH is $15.1\%$ larger than that of the dBH and the concentration has been reduced by a factor of $47\%$.   For the two component disk model, the $V_{200}$ of our PH is $19.8\%$ larger than that of the dBH and the concentration has been reduced by a factor of $61\%$. 

\subsubsection{NGC 3621}
\begin{figure}
\epsfig{figure=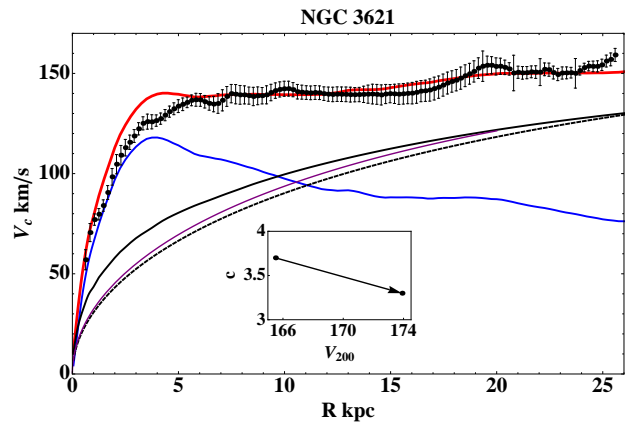,scale=0.37}
\caption{Rotation curve fit for NGC 3621.  Lines and inset are the same as in Figure \protect\ref{NGC24031}.}
\label{NGC3621} 
\end{figure}

NGC 3621 is a late-type spiral galaxy.  We use a one component disk to model the stellar component and see that the baryonic component is nearly maximal.  Our CH does well at $R>7$ kpc, fitting $V_{flat}$ almost perfectly (see Figure \ref{NGC3621}).  Despite this success, we find that our fit is not nearly as accurate as the dBH.  Because the baryonic component is nearly maximal at the inner radii, we find that in order to fit $V_{flat}$, we will over predict the rotation curve at the inner radii.  We can see that the CH has more mass than the dBH out to $R\sim20$ kpc.  We find that the $V_{200}$ of our PH is only $5.2\%$ larger than that of the dBH and the concentration has been reduced by a factor of $14\%$.  This scenario is easily resolvable by choosing a lower $M_*/L$.  

\subsubsection{NGC 4736}
NGC 4736 is one of the more unique galaxies in our sample, having a steeply declining rotation curve as well as a nearly maximum disk.  We use a two component model for the disk and scaled the mass of the bulge consistent with \cite{dB2008} so that the baryonic component would not be super-maximal.  The rotation curve out to all radii sampled is already well described by the baryonic component with little need for dark matter.  Furthermore, there are multiple features in the rotation curve and there is evidence of noncircular motion in the disk.

\begin{figure}
\epsfig{figure=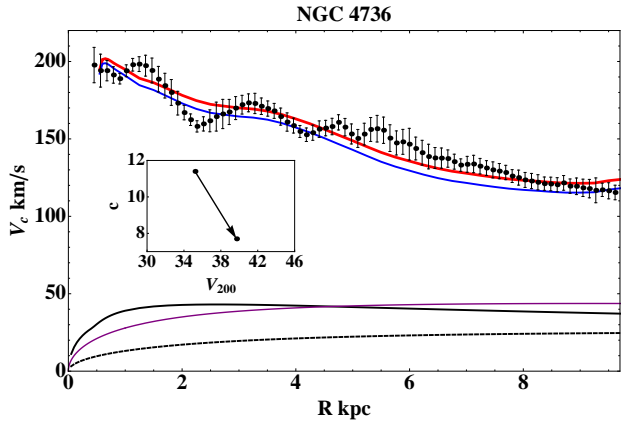,scale=0.37}
\caption{Rotation curve fit for NGC 4736.  Lines and inset are the same as in Figure \protect\ref{NGC24031}.}
\label{NGC4736} 
\end{figure}

Because the rotation curve is so well described by the baryons, this galaxy has the least significant dark matter halo of all the galaxies in our sample.  Despite this, our CH reveals a comparable fit to the dBH (see Figure \ref{NGC4736}).  The trend of a higher $V_{200}$ and smaller concentration is also evident however, there is much more freedom to choose halo parameters because the dark matter component relatively insignificant for this galaxy.

\subsubsection{DDO 154}
DDO 154 is a prime example of a well studied gas-rich dwarf galaxy.  Because of its gas-rich nature, changes to $M_*/L$ have little effect on the resulting best fit halo.  We use a one component disk and unlike NGC 4736, the dark matter component for this galaxy will dominate over the baryonic component for all radii.

\begin{figure}
\epsfig{figure=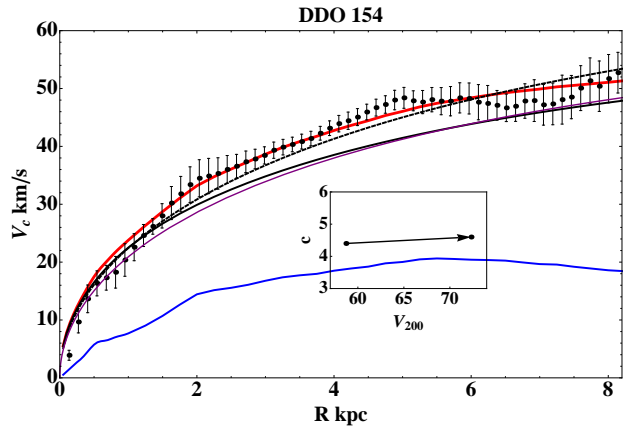,scale=0.37}
\caption{Rotation curve fit for DDO 154.  Lines and inset are the same as in Figure \protect\ref{NGC24031}.}
\label{DDO154} 
\end{figure}

Our CH does reasonably well fitting this galaxy (see Figure \ref{DDO154}).  It is interesting to note that the CH has been overtaken by the PH at very small radii, unlike all other galaxies in our sample where this occurs at radii larger than what has been sampled.  Because DDO 154 is so light, very little compression actually occurs.  Our algorithm conserves mass by assuming that the PH is a well mixed distribution of dark matter and baryons.  In order to form the disk, the baryons are then removed from the PH, resulting in a CH that is simply the remnant of PH without the baryonic component.  With this in mind, the $V_{200}$ of the PH is $28.7\%$ larger than that of the dBH, while the concentration has remained reasonably unchanged.

\subsubsection{NGC 7793}
NGC 7793 is a late-type Sd spiral galaxy.  We use a one component model for the disk and note that the rotation curve begins declining at $R\sim5$ kpc.  The degree of this decline is relatively uncertain and varies for different studies of this rotation curve.

\begin{figure}
\epsfig{figure=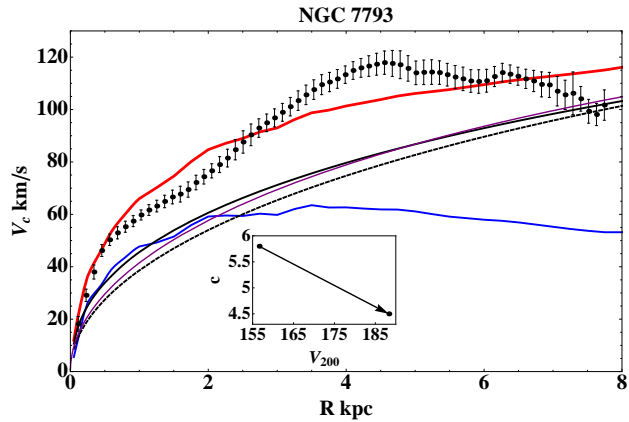,scale=0.37}
\caption{Rotation curve fit for NGC 7793.  Lines and inset are the same as in Figure \protect\ref{NGC24031}.}
\label{NGC7793} 
\end{figure}

Because of the declining rotation curve and the necessity for a dark matter component comparable to baryonic component at the inner radii of this galaxy, any NFW halo and even more so, our CH is doomed to failure for this galaxy.  The $\chi^2$ minimization algorithm will look to fit this galaxy by over predicting the velocity at large radii and under predicting the large bump at $3<R<5$ kpc and once again over predicting the velocity at smaller R (see Figure \ref{NGC7793}). The shape of the NFW halo is simply inconsistent with this galaxy.  Our CH has more trouble because more mass is concentrated in the center and we over predict the rotation curve more than an ordinary NFW halo fit.  

\subsection{Low Concentration Galaxies}
A clear trend among the fits thus far has been that the CH is more massive at the inner radii of the galaxies compared to the dBH.  When the concentrations of the dBH approach zero, the compression of any PH will, by construction, provide a worse fit to the rotation curve.  These types of galaxies often have super-maximal disks with a baryonic component that declines below $V_{flat}$ at smaller R.  In other words, the dark matter component must be minimal in the inner portion of the galaxy as to not significantly over predict the rotation curve, but increase steeply at larger R to make up for the decline in the baryonic component.  A halo of this shape is inconsistent with the shapes of the compressed haloes derived in the previous subsection. 

This problem is resolvable by decreasing $M_*/L$ and we provide examples where adopting a Kroupa IMF over the diet-Salpeter IMF will allow for reasonable compression without an unrealistically low concentration.  One might also note that Young's algorithm does not over predict compression when a dominant baryonic component is present and therefore, the calculated compression for these galaxies is significantly different than what would be predicted using the Blumenthal method.  For consistency, we do not include the fits to these galaxies in our further analysis of the 9 galaxies in the previous subsection.

It must also be considered that although adopting a Kroupa IMF may allow for a better fit for a compressed halo, this does not necessarily rule out the choice of a diet-Salpeter IMF for a specific galaxy.  There is an inherent scatter in $M_*/L$ when adopting a diet-Salpeter IMF for a galaxy of a given colour \citep{Bell2001}.  This scatter may cause the disk to become sub-maximal and thus potentially allowing for a reasonable fit using a compressed halo and a diet-Salpeter IMF.  Hence, adopting a stellar IMF with a lower $M_*/L$ is guaranteed to make the galaxy more susceptible to a compressed halo fit; however, it often cannot rule out a heavier IMF due to the inherent scatter in $M_*/L$.
Indeed, the small scatter in the Tully-Fisher relation is predominantly attributable to scatter in the mass-to-light ratio, and precludes arbitrary variation of the IMF from galaxy to galaxy \citep{V01,mcgaugh2005}.  Here, we adopt a Kroupa IMF for the LCGs for specificity in illustrating the effect of a different mass-to-light ratio.

\subsubsection{NGC 3521}
\begin{figure}
\epsfig{figure=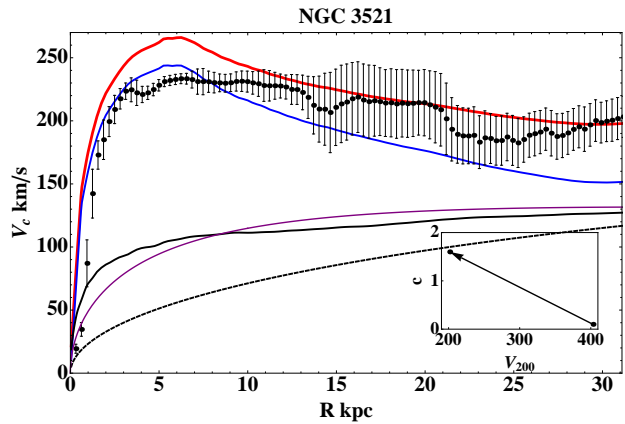,scale=0.37}
\caption{Rotation curve fit for NGC 3521.  Lines and inset are the same as in Figure \protect\ref{NGC24031}.}
\label{NGC3521a} 
\end{figure}

We use a one component disk model to describe the stellar component of the rotation curve.  The baryonic component rises slightly above observations from about $3<R<7$ kpc and then declines below observations at about $R>8$ kpc, resulting in the necessity for a significant dark matter halo.  In order to fit the outer portion of the rotation curve, we significantly over predict the inner portion due to the strong compression from having a dominant baryonic component (see Figure \ref{NGC3521a}).  

\begin{figure}
\epsfig{figure=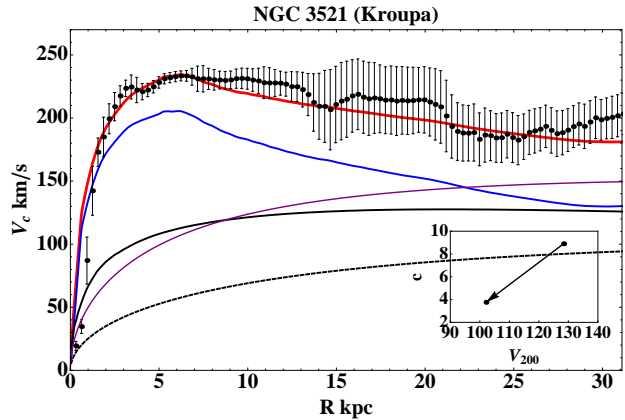,scale=0.37}
\caption{Rotation curve fit for NGC 3521 (Kroupa stellar IMF).  Lines and inset are the same as in Figure \protect\ref{NGC24031}.}
\label{NGC3521b} 
\end{figure}

When adopting the Kroupa IMF, the disk is no longer super maximal.  Our CH differs significantly from the dBH, but is statistically, a better fit (see Figure \ref{NGC3521b}).  The dBH has a much higher $V_{200}$ and concentration than the best fit PH.  Neither fit is able to match the inner few data points leading to a higher $\chi^2_{\nu}$ than other galaxies in the sample; however, the CH fit does very well at about $R>1.5$ kpc.

\subsubsection{NGC 5055}
\begin{figure}
\epsfig{figure=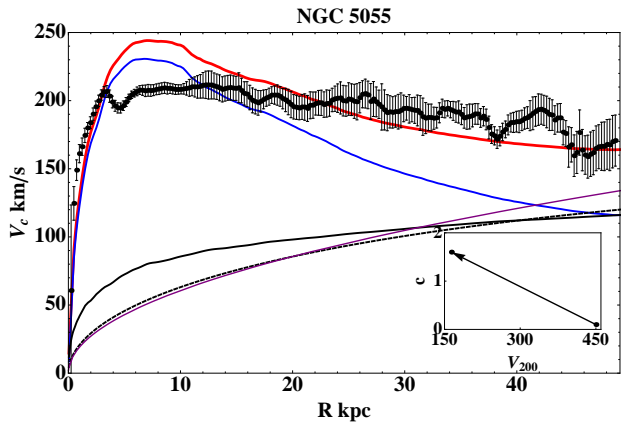,scale=0.37}
\caption{Rotation curve fit for NGC 5055.  Lines and inset are the same as in Figure \protect\ref{NGC24031}.}
\label{NGC5055a} 
\end{figure}

NGC 5055 is an Sbc galaxy and we use a two component disk to model the stellar contribution.  The $M_*/L$ for the bulge has been altered from what is predicted to match \cite{dB2008} and decrease the over prediction of the rotation curve at small radii.  Despite this effort, the baryonic component still significantly over predicts the rotation curve at about $4<R<12$ kpc.  The baryonic component declines even more steeply than in NGC 3521, once again requiring a significant dark matter contribution at larger radii.  In order to fit the outer portion of the rotation curve, we significantly over predict the inner portion due to the strong compression from having a dominant baryonic component in the inner region (see Figure \ref{NGC5055a}).

\begin{figure}
\epsfig{figure=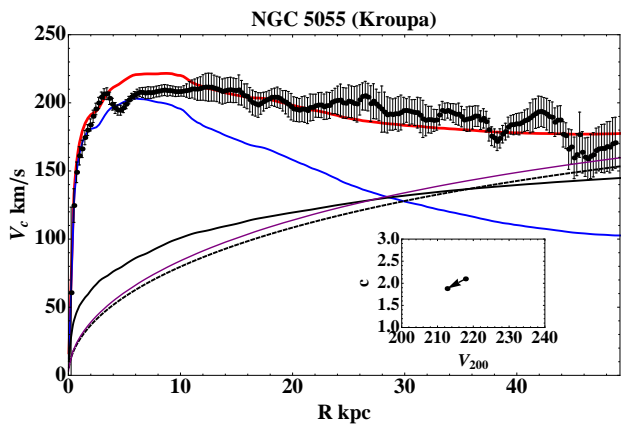,scale=0.37}
\caption{Rotation curve fit for NGC 5055 (Kroupa stellar IMF).  Lines and inset are the same as in Figure \protect\ref{NGC24031}.}
\label{NGC5055b} 
\end{figure}

When adopting the Kroupa IMF, the disk is no longer over predicts the rotation curve, but is maximal.  Our CH does significantly worse than the dBH for this galaxy because of the presence of a significant decline in the baryonic component as well as a maximal disk (see Figure \ref{NGC5055b}).  As was stated for the diet-Salpeter fit for this galaxy, in order to fit the outer portion of the rotation curve, we significantly over predict the inner portion due to the strong compression from having a dominant baryonic component in the inner region.

\subsubsection{NGC 7331}
\begin{figure}
\epsfig{figure=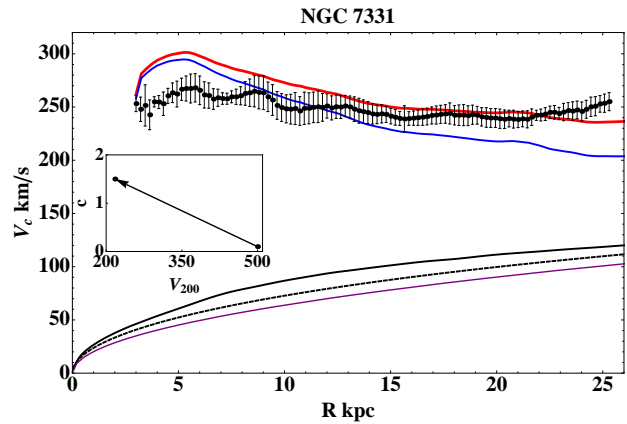,scale=0.37}
\caption{Rotation curve fit for NGC 7331.  Lines and inset are the same as in Figure \protect\ref{NGC24031}.}
\label{NGC7331a} 
\end{figure}

NGC 7331 is a late-type Sd spiral galaxy with a significant colour gradient.  We use a two component disk to model the stellar contribution.  In order to be consistent with \cite{dB2008}, we fit the rotation curve without a colour gradient due to the incompatibility of their mass models with the colour gradient (see Figure \ref{NGC7331a}).  As is common among these low concentration galaxies, the baryonic component is super-maximal, and over predicts the rotation curve at about $3<R<8$ kpc.  Because there is only a shallow decline in the baryonic component, this galaxy requires respectively less dark matter than the other two in this subsection.  Hence, our measure for $\chi_{\nu}^2$ is more comparable to that of the dBH.  

\begin{figure}
\epsfig{figure=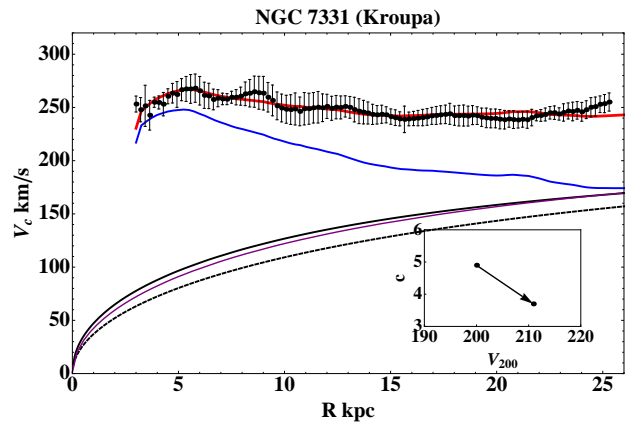,scale=0.37}
\caption{Rotation curve fit for NGC 7331 (Kroupa stellar IMF).  Lines and inset are the same as in Figure \protect\ref{NGC24031}.}
\label{NGC7331b} 
\end{figure}

When adopting the Kroupa IMF, the disk is no longer super maximal and our CH provides a very good fit for this galaxy, comparable to the dBH (see Figure \ref{NGC7331b}).  Despite the disk being nearly maximal for this galaxy, the baryonic contribution to the rotation curve declines slowly; thus requiring a less dominant dark matter component at outer radii.  Contrary to NGC 5055, the nearly maximal disk does not over compress the dark matter at the inner regions, allowing for a very good fit.

\section{Discussion}
\begin{figure}
\epsfig{figure=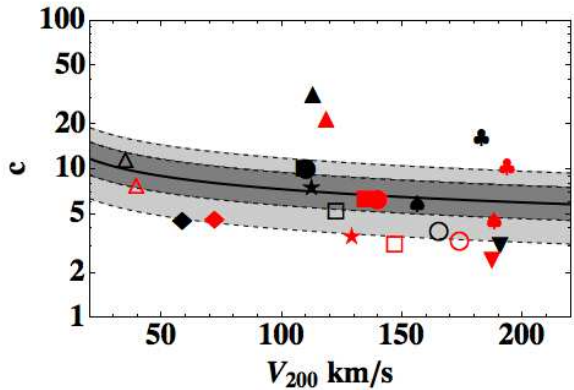,scale=0.4,trim= 25 200 0 200}
\caption{The c-$V_{200}$ relationship for our primordial NFW haloes (red symbols) as well as for the \protect\cite{dB2008} NFW haloes (black symbols).  Corresponding symbols represent the same galaxy which can be identified in Table 1.  The gray line represents the expected c-$V_{200}$ relation from $\Lambda$CDM as derived in \protect\cite{Maccio2008}.}
\label{v200c} 
\end{figure}

Having accurately computed halo compression for multiple different types of galaxies, we now aim to connect the observed properties of these galaxies with their host dark matter halo; bridging the gap between dark matter N-Body simulations and disk galaxy formation.  In Figure \ref{v200c} we plot the NFW halo parameters from the \cite{dB2008} NFW halo fits along with the parameters of the primordial haloes from our fits and compare with predictions from $\Lambda$CDM \citep{Maccio2008}.  The halo parameters derived from our compression algorithm broadly agree with the $c-V_{200}$ relation predicted by $\Lambda$CDM, though more galaxies lie on the
low concentration side of the relation.  We note that the NFW halo fits from \cite{dB2008} also tend to agree with the $c-V_{200}$ relation predicted by $\Lambda$CDM.  Since the primordial NFW haloes parameters calculated from our compression algorithm tend to lie at the lower end of the concentration region predicted by $\Lambda$CDM, one may argue that this work may favor the notion that actual haloes either exhibit no compression or possibly mild expansion.  This statement is restricted to the galaxies we have been able to fit here, and does not address the broader problem of $\Lambda$CDM over-predicting the mass in the centers of low mass and low surface brightness galaxies \citep{KZN2008,KZN2009,dB2010}.  There are galaxies within the THINGS sample that we were unable to fit with our compression algorithm which are likewise inconsistent with an ordinary NFW halo fit as well (these include NGC 925, NGC 2366, NGC 2976, and IC 2574).  Competing effects such as dynamical friction and outflows counteract compression and both processes likely affect real galaxies. In low surface brightness galaxies in particular, the density profile requires halo expansion \citep{Oh2011}.  However, there are certainly cases in our small sample where the primordial halo parameters derived from our compression algorithm agrees with the $c-V_{200}$ relation predicted by $\Lambda$CDM better than the current NFW halo fit.  This suggests that compression may be the dominant effect in higher mass galaxies, but it may be that there is no unique approach to modeling the density profile.

The general trend is that compression leads primordial haloes to have a higher $V_{200}$ and lower concentration than what is inferred from directly fitting an NFW halo to the rotation curve.  Estimates of these parameters by directly fitting NFW haloes will systematically overestimate the true concentration and under estimate the true virial velocity of the primordial halo as predicted by dark matter simulations. 

\begin{figure*}
  \centering
  \begin{tabular}{cc}
    \epsfig{figure=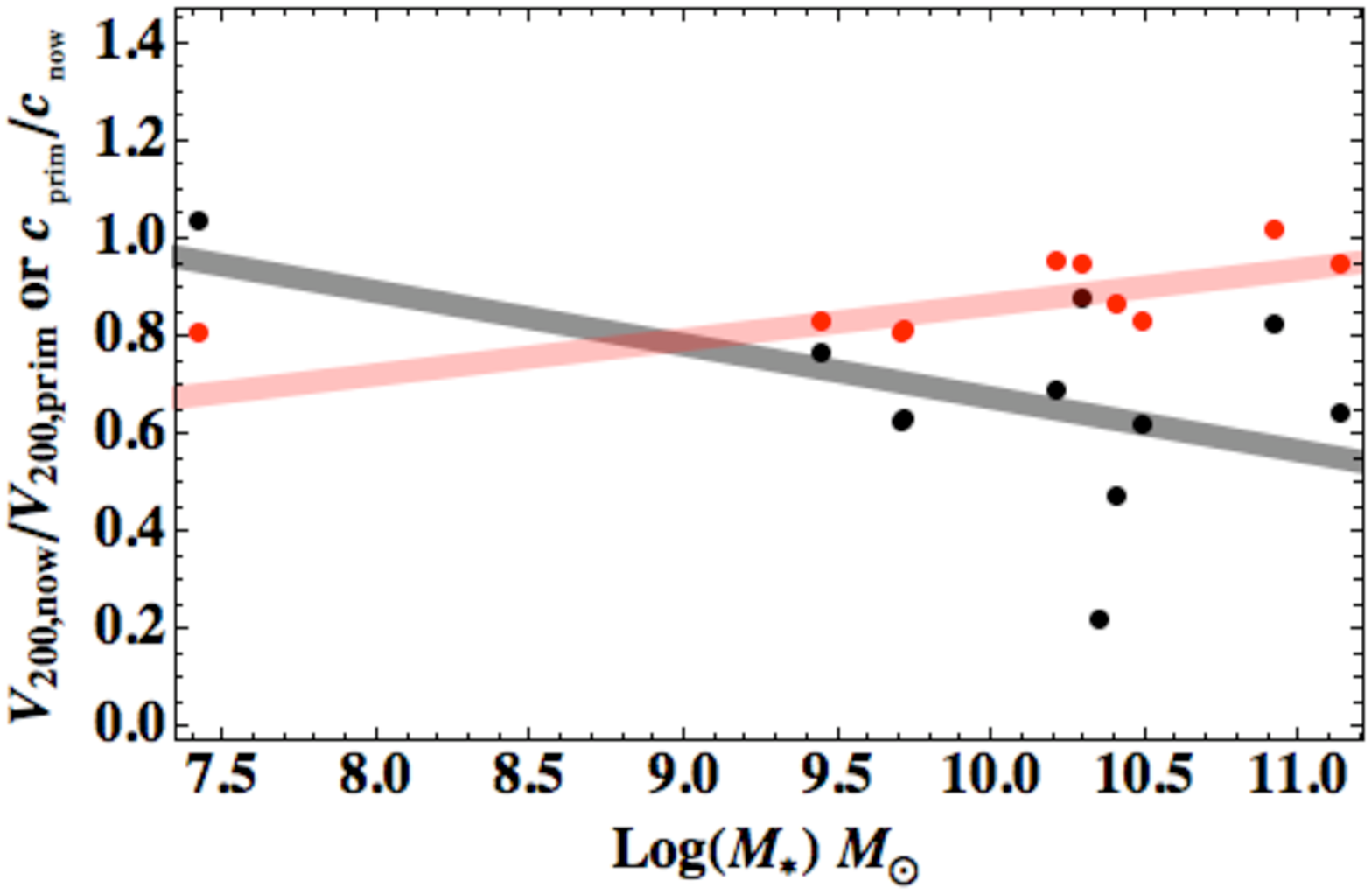,scale=0.42,trim= 25 200 0 200}&
    \epsfig{figure=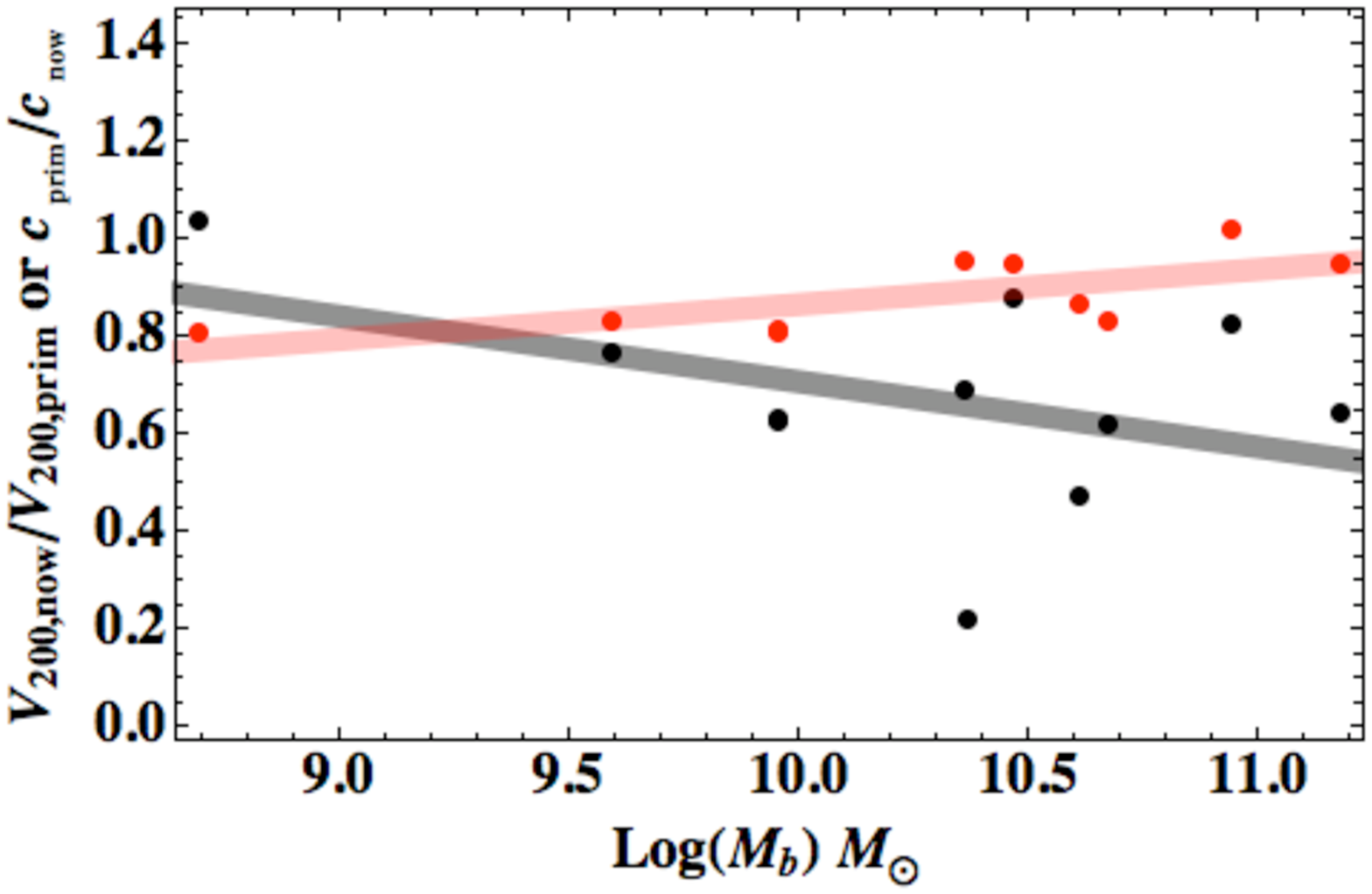,scale=0.42,trim= 25 200 0 200}\\
  \end{tabular}
  \caption{{\it Left.} Ratios of our primordial NFW parameters with the \protect\cite{dB2008} NFW halo parameters compared to the stellar mass of the galaxy.  {\it Right.} Ratios of our primordial NFW parameters with the \protect\cite{dB2008} NFW halo parameters compared to the baryonic mass (stars+gas) of the galaxy.  In both plots, red points represent the $V_{200,now}/V_{200,prim}$ and black points represent $c_{prim}/c_{now}$.  The red line and black lines are the respective best fits.  The data for NGC 4736 has been excluded for its vast inconsistency with the rest of the sample.}
  \label{ratiofit}
\end{figure*}

We can use the apparent relation between compression and mass (Fig.~\ref{ratiofit}) to prescribe an analytic method to map from a directly fitted NFW halo to the corresponding primordial halo.  One simply fits the rotation curve with an NFW halo and uses the following relations to map the direct fit NFW halo parameters to the primordial NFW halo parameters:

\begin{equation}
V_{200,prim}(M_b)=\frac{V_{200,fit}}{A_V+B_V\log(M)}
\end{equation}
\begin{equation}
c_{prim}(M_b)=c_{fit}(A_c+B_c\log(M))
\end{equation}
Where M is either $M_b$ or $M_*$ (The fit constants take different values when choosing $M_b$ or $M_*$).

In Figure \ref{ratiofit}, we plot the ratios of our primordial halo NFW halo parameters with the best fit \cite{dB2008} parameters against the baryonic mass and stellar mass to determine the constants in these equations.  We find: $A_V=0.145$, $B_V=0.072$, $A_c=2.047$ and $B_c=-0.134$ when relating to $M_b$ and $A_V=0.417$, $B_V=0.047$, $A_c=1.762$ and $B_c=-0.109$ when relating to $M_*$.  We emphasize that there is significant error associated with these values due to our small sample of galaxies as well as scatter in the data.  The $V_{200}$ relation exhibits a much tighter trend compared with the concentration relation where there is substantial scatter.  There is also a large gap between most galaxies in our sample at the higher mass end and the sole galaxy, DDO 154, in the low mass regime.  In order to further constrain the fit constants, future studies would need to fill this gap.

\begin{figure}
\epsfig{figure=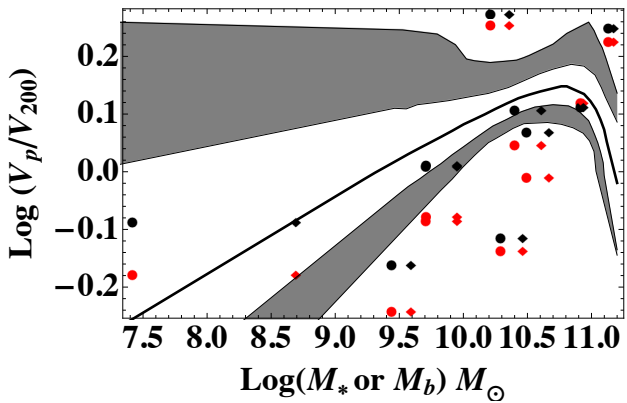,scale=.69,trim= 180 200 200 220,angle=90}
\caption{Relationship between the stellar mass and baryonic mass with $V_{p}/V_{200}$ where $V_p$ is the circular velocity at the radius where the baryonic mass model peaks. This is equivalent to $V_{2.2}$ for purely exponential discs \citep{McGaugh2005b}.  Black points represent value from the \protect\cite{dB2008} NFW halo fits and red points represent the value from our primordial NFW haloes.  The circles are plotted against only the stellar mass whereas the diamonds are plotted against the total baryonic mass (stars $+$ gas).  The shaded region is the expectation from \protect\cite{Reyes2012} for only the stellar mass relation.  The $1\sigma$ confidence region in white and the $2\sigma$ confidence region shown in gray.  The regions have been corrected to reflect the stellar IMF used in this paper.}
\label{r2} 
\end{figure}

This relation can also be used to relate haloes produced in simulations to a galaxy it will likely contain.  Simply reverse the relation by inputting the primordial halo parameters and assuming that $M_b=f_bf_dM_{200}$, where $f_b$ is the cosmic baryon fraction, $M_{200}$ is the mass of the halo, and $f_d$ is the fraction of the baryons that contribute to the observed galaxy.  Constraints from the baryonic Tully-Fisher relation (BTFR) \citep{McGaugh2012} 
suggest\footnote{Note that the precise value of the constant in this equation depends on the choice of $\Delta = 200$, the calibration of the BTFR, 
and mildly on cosmological parameters --- see \citet{McGaugh2012}.} that:
\begin{equation}
\log(f_d)= \log(\frac{f_v^3 V_{f}}{100\ km\ s^{-1}}) -1
\end{equation}
where $f_v=\frac{V_{flat}}{V_{vir}}$.  The choice of $f_v$ is not particularly obvious.  Using $V_{2.2}$ rather than $V_{flat}$, \cite{Reyes2012} found $1\leq f_v\leq1.3$ for high mass galaxies (which is similar to earlier studies by \citealt{Eke2006} $\&$ \citealt{Dutton2010}); however there is substantial scatter around these values (see Figure \ref{r2} for how this value relates to the stellar mass of the galaxy).  It is quite possible that the relation between $M_h$ and $M_b$ is not one-to-one and thus there is some range in possible choices of $f_d$.  This range must be modest in order to not induce a large amount of scatter in the Tully-Fisher relation, or there must be some fine-tuned covariance between $f_d$ and other parameters \citep{McGaugh2010}

\begin{figure}
\epsfig{figure=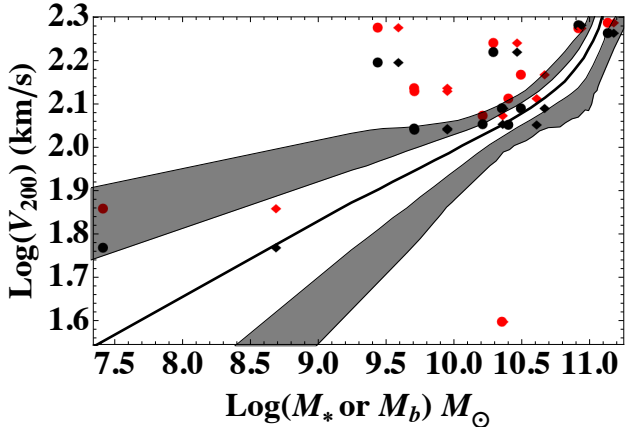,scale=.69,trim= 180 200 200 220,angle=90}
\caption{Relationship between the stellar mass and baryonic mass with $V_{200}$.  Black points represent value from the \protect\cite{dB2008} NFW halo fits and red points represent the value from our primordial NFW haloes.  The circles are plotted against only the stellar mass whereas the diamonds are plotted against the total baryonic mass (stars $+$ gas).  The shaded region is the expectation from \protect\cite{Reyes2012} for only the stellar mass relation.  The $1\sigma$ confidence region in white and the $2\sigma$ confidence region shown in gray.  The regions have been corrected to reflect the stellar IMF used in this paper.}
\label{r1} 
\end{figure}

\begin{figure}
\epsfig{figure=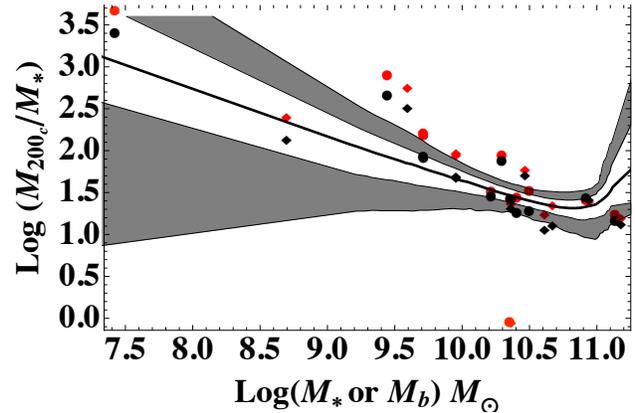,scale=.69,trim= 180 200 200 220,angle=90}
\caption{Relationship between the stellar mass and baryonic mass with $M_{200}/$$(M_{*}$ or $M_b)$.  Black points represent value from the \protect\cite{dB2008} NFW halo fits and red points represent the value from our primordial NFW haloes.  The circles are plotted against only the stellar mass whereas the diamonds are plotted against the total baryonic mass (stars $+$ gas).  The shaded region is the expectation from \protect\cite{Reyes2012} for only the stellar mass relation.  The $1\sigma$ confidence region in white and the $2\sigma$ confidence region shown in gray.  The regions have been corrected to reflect the stellar IMF used in this paper}
\label{r3} 
\end{figure}

Continuing with the idea of relating the primordial NFW parameters with properties of the constituent spiral galaxy, we compare our primordial halo parameters with observational constraints on the observed mass-$V_{200}$ relation (OMVR) as well as the halo-to-stellar mass relation (HSMR) and the optical-to-virial velocity relation (OVVR) from \cite{Reyes2012} (see Figures \ref{r1}-\ref{r3}).  We can see in all three figures that our results are broadly consistent.  However, there is considerable scatter, and important differences in detail.  

Figures \ref{r1}-\ref{r3} show both the NFW fit to the current dark matter distribution \citep{dB2008} and the primordial halo before compression.  \cite{Reyes2012} employ a somewhat lighter IMF than we do here.  This would place the corresponding primordial halo intermediate between our result and that of \cite{dB2008}, which is  the limit of no compression.  The difference, while apparent, is modest.  It cannot reconcile the discrepant cases that fall outside the band of \cite{Reyes2012}.  The effect is not large enough for individual galaxies, and a systematic change to the IMF would improve the agreement of some cases only at the expense of making those currently in good agreement worse.  It is likely that the population models simply misestimate the mass-to-light ratio in some cases, but these are not necessarily the discrepant cases.

Taking our own result at face value, $V_p \approx V_{200}$ with substantial scatter ($\sim 0.2$ dex: Figure \ref{r2}).  We infer similar or somewhat larger $V_{200}$ at a given stellar mass than do \cite{Reyes2012}, and a smaller $f_v$ (1.0 rather than 1.3).  This in itself is not inconsistent with the OVVR of \cite{Reyes2012} as we have not averaged over the scatter in a similar fashion.  Clearly the observed velocity is related to that of the parent dark matter halo, but considerable caution must be exercised in relating one to the other for any given galaxy.

There is a hint that our data follow the same trend with mass as suggested by the bands in Figures \ref{r1}-\ref{r3}, but there is not enough data to say anything definite.  Indeed, the information at low masses from all sources is rather patchy and not obviously trustworthy.  We can neither confirm nor refute the trend in $f_v$ inferred from the halo occupation distribution \citep{More2011} or the Tully-Fisher relation \citep{McGaugh2012}.  The NFW halo model rarely fits low mass galaxies, so it is not obvious that extrapolations to low mass can be meaningful. 

\section{Conclusion}
We have modeled the primordial haloes of 12 galaxies from the THINGS survey by using Young's algorithm to exactly calculate adiabatic compressions of dark matter haloes.  We find a general trend that concentration and virial velocity of the primordial halo decrease and increase respectively when compared to the best fit NFW halo of the rotation curve.  For the majority of galaxies, we find that the compressed halo is statistically comparable to the regular best fit NFW halo and in some cases even better.  


The NFW parameters for the primordial haloes fall on and slightly below the c-$V_{200}$ relation predicted by $\Lambda$CDM.  It appears possible that adiabatic contraction is dominant effect that reshapes the dark matter haloes of massive spiral galaxies. This does not preclude the influence of other effects, like feedback.  Indeed, such a process does appear to be necessary in some galaxies, but not necessarily in all.  It is tempting to conclude that feedback becomes progressively more relevant in lower mass galaxies, but there can be exceptions at both high and low mass, so one must be careful not to over-generalize.

A trend has emerged between the virial velocity of the primordial halo and the stellar/baryonic mass of a galaxy.  A similar trend is also present for the concentration of the primordial halo and the galaxy mass, albeit with greater scatter.  The relations derived from these trends can be used to estimate the parameters of the primordial halo of observed galaxies without formally computing the compression.  A larger sample of galaxies would be useful to better constrain these relations.

\section{Acknowledgements}
HK's work is partially supported by Foundation Boustany, Cambridge Overseas Trust, and the Isaac Newton Studentship.  This work has been supported in part by NSF grant AST 0908370 to S.M.\ and AST 1211793 to J.S.  WJGdB was supported by the European Commission (grant FP7-PEOPLE-2012-CIG \#333939).  We thank the referee for some of the most insightful and useful suggestions that the authors have received in their joint experience.

\bibliographystyle{./apj}
\bibliography{./KMSdB2014}

\label{lastpage}
\end{document}